\documentclass[11pt,epsf]{article}
\usepackage{comment}
\newcommand {\dfn} {\stackrel{\Delta} {=}}

\newcommand{\gea}{\stackrel{\mbox{(a)}}{\ge}}
\newcommand{\geb}{\stackrel{\mbox{(b)}}{\ge}}
\newcommand{\gec}{\stackrel{\mbox{(c)}}{\ge}}

\newcommand {\bu} {\mbox{\boldmath $u$}}

\newcommand {\bx} {\mbox{\boldmath $x$}}
\newcommand {\by} {\mbox{\boldmath $y$}}
\newcommand {\bz} {\mbox{\boldmath $z$}}

\newcommand{\calA}{{\cal A}}

\newcommand{\calE}{{\cal E}}

\newcommand{\calL}{{\cal L}}

\newcommand{\calN}{{\cal N}}

\newcommand{\calP}{{\cal P}}

\newcommand{\calT}{{\cal T}}
\newcommand{\calU}{{\cal U}}

\newcommand{\calX}{{\cal X}}
\newcommand{\calY}{{\cal Y}}
\newcommand{\calZ}{{\cal Z}}
\newcommand{\hP}{\hat{P}}

\begin{document}
\thispagestyle{empty}
\title{Universal Encryption of Individual\\
 Sequences Under Maximal Leakage
}
\author{Neri Merhav
}
\date{}
\maketitle

\begin{center}
The Andrew \& Erna Viterbi Faculty of Electrical Engineering\\
Technion - Israel Institute of Technology \\
Technion City, Haifa 32000, ISRAEL \\
E--mail: {\tt merhav@ee.technion.ac.il}\\
\end{center}
\vspace{1.5\baselineskip}
\setlength{\baselineskip}{1.5\baselineskip}

\begin{abstract}
We consider the Shannon cipher system in the framework of individual sequences
and finite-state encrypters under the metric of maximal leakage
of information. A lower bound and an asymptotically matching upper bound on
the leakage are derived, which lead to the conclusion that asymptotically
minimum leakage can be attained by Lempel-Ziv compression followed by one-time pad
encryption of the compressed bit-stream.
\end{abstract}

\setcounter{section}{0}

\section{Introduction}
\label{intro}

Theoretical frameworks centered on the combination of individual sequences and 
finite-state encoders and decoders, have been thoroughly explored, marking a significant departure 
from the traditional probabilistic models typically employed in source and channel modeling. 
This shift has been particularly noticeable in a variety of information-theoretic fields, 
including data compression \cite{KY93}, \cite{MZ06}, \cite{RM11}, \cite{WMF94}, \cite{WM02}, 
\cite{YK96}, \cite{Ziv78}, \cite{Ziv80}, \cite{Ziv84}, \cite{ZL78}, 
source/channel simulation \cite{MMSW10}, \cite{Seroussi06}, 
classification \cite{Merhav00}, \cite{Ziv88}, \cite{ZM93}, prediction \cite{FMG92}, \cite{HKW98}, 
\cite{MF93}, \cite{WM01}, \cite{WMSB01}, \cite{ZM07}, 
denoising \cite{WOSVW05}, and even channel coding \cite{LF11}, \cite{LF13}, \cite{SF05}. 
For a concise recent overview, see \cite{mebits}. These references only skim the surface of a much larger body of work.
In sharp contrast, the field of information-theoretic security, 
from Shannon's pioneering work \cite{Shannon48} to more recent studies 
\cite{Hellman77}, \cite{Lempel79}, \cite{LPS09}, \cite{Massey88}, \cite{Yamamoto91}, 
has remained almost entirely rooted in the probabilistic framework. Although these examples represent 
just a small fraction of the extensive literature, they highlight the near-exclusive reliance on probabilistic models within this domain.

Only two notable exceptions to this prevailing paradigm are known
to the author: one is an
unpublished memorandum by Ziv \cite{Ziv78u} and the other is a subsequent work \cite{me13}. 
Ziv’s memorandum introduces a distinctive approach where the plaintext source, intended for encryption with a secret key, is treated as an 
individual sequence. In this model, the encrypter is conceptualized as a general block encoder, while the eavesdropper employs a 
finite-state machine (FSM) to distinguish between messages. Ziv hypothesizes that the eavesdropper has some prior knowledge of the plaintext, 
which is expressed as a set of ``acceptable messages,'' referred to as the acceptance set. In other words, 
prior to observing the ciphertext, the eavesdropper's uncertainty about the plaintext is that it could be any 
member of this set of acceptable messages.
According to Ziv’s framework, perfectly secure encryption occurs when the presence of the ciphertext does not reduce the 
uncertainty about the acceptance set. In essence, even after intercepting the ciphertext, the eavesdropper 
learns nothing new about the plaintext that she did not already know. The size of the acceptance set serves as a measure of uncertainty: 
a larger set corresponds to greater uncertainty. The FSM is then used to distinguish between acceptable and unacceptable 
plaintext sequences based on various key bit sequences. Consequently, perfect security 
is defined as maintaining the size of the acceptance set, and thus the uncertainty, unchanged in the presence of the ciphertext.
Ziv’s primary finding is that the asymptotic key rate necessary for perfectly secure encryption, 
according to this definition, cannot be lower (up to asymptotically vanishing terms) than the Lempel-Ziv (LZ) complexity 
of the plaintext source \cite{ZL78}. Notably, this lower bound can be asymptotically achieved using one-time pad encryption 
(i.e., bit-by-bit XOR with key bits) on the bit-stream generated by LZ data compression of the plaintext, 
echoing Shannon's classical probabilistic result that the minimum key rate
required is equal to the source's entropy rate. More recently, Ziv’s
methodology has been refined and expanded in several directions in \cite{me23}.

In the follow-up work \cite{me13}, the concept of perfect secrecy 
for individual sequences was approached from a different perspective. 
Rather than assuming a finite-state eavesdropper with 
predefined knowledge, this framework posits that the encrypter itself can be modeled 
as a FSM, which is sequentially fed both the plaintext source and random key
bits. A new concept, ``finite-state encryptability'', is introduced, inspired by the analogous idea of 
finite-state compressibility in \cite{ZL78}. This concept defines the minimum key rate that must be used by any 
finite-state encrypter to ensure that a certain form of normalized empirical mutual information 
between the plaintext and ciphertext tends to zero as the block length grows.
Among the key results in \cite{me13}, it is established and proven that the finite-state encryptability of an individual sequence 
is fundamentally bounded from below by its finite-state compressibility. This lower bound is again asymptotically achieved by 
applying LZ compression to the plaintext and then one-time pad encryption of
the compressed bits.

In this paper, we adopt the same model setting as in \cite{me13}, but with a different security metric: 
the maximum leakage of information, which was first introduced by Issa, Wagner, 
and Kamath in \cite{IWK20} and then further explored in several more
recent works, 
including \cite{BGYOPSS21}, \cite{EGI21}, \cite{KSK24}, \cite{SCOS23}, 
and \cite{WBZ25}, among others. This metric is closely related to, 
and similarly motivated by, the earlier security measure proposed 
in \cite{me03}, which defines security as a scenario where the correct 
decoding exponent of the plaintext is not improved by the 
availability of the ciphertext, compared to that of blind guessing. For more
details, see the last paragraph of Subsection \ref{mil}.
The maximum leakage metric is defined in a more general form and has a relatively 
straightforward expression, as demonstrated in \cite{IWK20} 
and further clarified in the following sections. As will be discussed in the
sequel, the maximum leakage metric is particularly well-suited for the individual-sequence setting considered here, 
as it is weakly dependent on the probability distribution of the plaintext,
depending only on its support.

We derive both a lower bound and an asymptotically matching upper bound on the leakage, 
leading yet again to the conclusion that asymptotically optimal performance can be achieved by applying 
LZ compression followed by one-time pad encryption of the compressed
bit-stream, and so, considering also the above mentioned earlier works,
\cite{Ziv78u}, \cite{me13}, and \cite{me23},
one of the messages of this work is that one-time pad
encryption on top of LZ compression forms an asymptotically optimal cipher system
from many aspects. That said, we believe that the deeper and more interesting
contribution of this work is the converse theorem (Theorem 1 in the sequel)
and its proof, asserting that the key rate that must be consumed to encrypt an
individual sequence cannot be much
smaller than the LZ complexity of the sequence minus the allowed normalized
maximal information leakage. 

The outline of the remaining part of this paper is as follows.
In Section \ref{ncbpf}, we establish notation conventions, provide some
necessary background, and formulate the problem studied in this work.
In Section \ref{main}, we assert the main results and discuss them.
Finally, in Section \ref{proof} we prove Theorem 1, which is the converse
theorem.

\section{Notation Conventions, Background and, Problem Formulation}
\label{ncbpf}

\subsection{Notation Conventions}
\label{nc}

Throughout this paper, scalar random
variables (RV's) will be denoted by capital
letters, their sample values will be denoted by
the respective lower case letters, and their alphabets will be denoted
by the respective calligraphic letters.
A similar convention will apply to
random vectors and their sample values,
which will be denoted with same symbols superscripted by the dimension.
Thus, for example, $A^m$ ($m$ -- positive integer)
will denote a random $m$-vector $(A_1,...,A_m)$,
and $a^m=(a_1,...,a_m)$ is a specific vector value in $\calA^m$,
the $m$--th Cartesian power of $\calA$. The
notations $a_i^j$ and $A_i^j$, where $i$
and $j$ are integers and $i\le j$, will designate segments $(a_i,\ldots,a_j)$
and $(A_i,\ldots,A_j)$, respectively,
where for $i=1$, the subscript will be omitted (as above).
For $i > j$, $a_i^j$ (or $A_i^j$) will be understood as the null string.
The notation $[u]_+$ for a real $u$ will stand for $\max\{0,u\}$.
Logarithms and exponents, throughout this paper, will be understood to be taken to the base 2
unless specified otherwise.

Sources and channels will be denoted generically by the letter $P$ or $Q$,
subscripted by the name of the RV and its conditioning,
if applicable, exactly like in
ordinary textbook notation standards, e.g., $P_{X^m}(x^m)$ is the probability function of
$X^m$ at the point $X^m=x^m$, $P_{X|W^m}(x|w^m)$
is the conditional probability of $X=x$ given $W^m=w^m$, and so on.
Whenever clear from the context, these subscripts will be omitted.
Information theoretic quantities, like entropies and mutual
informations, will be denoted following the usual conventions
of the information theory literature, e.g., $H(K^m)$, $I(V;X^m|W^m)$,
and so on.

In the sequel $x^n=(x_1,\ldots,x_n)$ will designate an individual sequence to
be encrypted. The components, $\{x_i\}$ of $x^n$ all take values in a finite
alphabet, $\calX$, whose cardinality will be denoted by $\alpha$.

\subsection{Background}
\label{bg}

Before the exposition of the main results and their proofs,
we revisit key terms and details
related to the notion of maximal leakage of information and the
1978 version of the LZ algorithm, also known as the LZ78
algorithm \cite{ZL78}, which is the central building block in this work.

\subsubsection{Maximal Leakage of Information}
\label{mil}

As mentioned in the Introduction, in this paper, we adopt the maximal leakage
\cite{IWK20} as our secrecy metric. For a probabilistic plaintext source, 
the maximal leakage from a secret random variable $X$, distributed according to
$\{P_X(x),~x\in\calX\}$, to another random variable $Y$, available to an
adversary, and which is conditionally distributed
given $X=x$ according to $\{P_{Y|X}(y|x),~x\in\calX,~y\in\calY\}$, is
defined as 
\begin{equation}
\label{def}
\calL(X\to Y)\dfn\sup_{U-X-Y-\hat{U}}
\log\frac{\mbox{Pr}\{\hat{U}=U\}}{\max_{u\in\calU} P_U(u)},
\end{equation}
where the supremum is over all finite-alphabet random variables $U$ and $\hat{U}$,
with the Markov structure $U-X-Y-\hat{U}$.
In other words, it is the maximum possible difference between the
logarithm of the probability of correctly guessing some (possibly randomized)
function of $X$ based on $Y$ and correctly guessing it blindly. 

In Theorem 1 of \cite{IWK20}, it was asserted and proved that the leakage can be
calculated relatively easily using the formula:
\begin{equation}
\calL(X\to
Y)=\log\left[\sum_{y\in\calY}\max_{\{x:~P_X(x)>0\}}P_{Y|X}(y|x)\right].\nonumber
\end{equation}
Clearly, if
$P_{Y|X}(y|x)$ is independent of $x$ for all $y\in\calY$ then $\calL(X\to Y)=0$,
which is the case of perfect secrecy. In general, the smaller is $\calL(X\to
Y)$, the more secure the system is.
In \cite{IWK20}, it is shown that the maximal leakage has many interesting
properties, one of them is that it satisfies a data processing inequality
(see Lemma 1 of \cite{IWK20}). It is also shown in Section III of \cite{IWK20} that the maximal leakage
has several additional operative meanings in addition to the original one
explained above.

Note that the dependence on the distribution of the secret random variable,
$P_X$, is rather weak, as it depends only on its support. When passing from
single variables to vectors of length $n$, $\calL(X^n\to Y^n)$ is defined in
the same manner except that $x$, $y$, $\calX$, $\calY$, $P_X(\cdot)$, and
$P_{Y|X}(\cdot|\cdot)$  are replaced by 
$x^n$, $y^n$, $\calX^n$, $\calY^n$, $P_{X^n}(\cdot)$, and
$P_{Y^n|X^n}(\cdot|\cdot)$, respectively. In this case, the weak dependence
of $\calL(X^n\to Y^n)$ on $P_{X^n}$ makes it natural to use when $P_{X^n}$ is
uncertain, or completely unknown, or even non-existent, such as in the
individual sequence setting considered here. In this case, we adopt the
simple definition
\begin{equation}
\label{ilformula}
\calL(x^n\to
Y^n)\dfn\log\left[\sum_{y^n\in\calY^n}\max_{x^n\in\calX^n}P_{Y^n|X^n}(y^n|x^n)\right],
\end{equation}
corresponding to the full support $\calX^n$ for $x^n$, which accounts for a
worst-case approach.
The operational significance of maximal information leakage in the
our setting can then be understood in two ways: (i)
Considering the definition (\ref{def}), it allows arbitrary probability
distributions (without any assumed structure) on $x^n$, including those that put almost all their mass on a single 
(unknown) arbitrary sequence, in the spirit of the individual-sequence setting
considered here. (ii) Referring to the formula (\ref{ilformula}), it is evident that the leakage 
vanishes whenever $P_{Y^n|X^n}(y^n|x^n)$ is independent of $x^n$, which is an
indisputable characterization for perfect secrecy in the individual-sequence
setting too, where no distribution at all is assumed on $x^n$. 

As mentioned in the Introduction, in \cite{me03} a somewhat
different security metric was proposed, but it is
intimately related to the maximal information leakage considered here.
In \cite{me03}, the idea was to define a system as secure if the probability
of guessing $X$ correctly is essentially the same if $Y$ is present or
absent. (More precisely, if $X$ and $Y$ are random vectors of dimension $n$, then a
system is considered secure if the correct decoding exponent of $X$ in the
presence of $Y$ is the same as if $Y$ is absent.)
Specifically, the correct decoding probability of $X$ based on $Y$ is
\begin{equation}
P_{\mbox{\tiny c}}=\sum_y\max_x P_{XY}(x,y),
\end{equation}
which is closely related to
\begin{eqnarray}
2^{\calL(X\to Y)}
&=&\sum_y\max_x P_{Y|X}(y|x)\nonumber\\
&=&|\calX|\cdot\sum_y\max_x \frac{P_{Y|X}(y|x)}{|\calX|}\nonumber\\
&\dfn&|\calX|\cdot\sum_y\max_x
P_X(x)P_{Y|X}(y|x)\nonumber\\
&=&\frac{\sum_y\max_x P_X(x)P_{Y|X}(y|x)}{1/|\calX|}\nonumber\\
&=&\frac{P_{\mbox{\tiny c}}^{\mbox{\tiny i}}}{P_{\mbox{\tiny
c}}^{\mbox{\tiny u}}},
\end{eqnarray}
where $P_X(\cdot)$ is understood to designate the uniform distribution across
$\calX$, and accordingly,
$P_{\mbox{\tiny c}}^{\mbox{\tiny i}}$ stands for the probability of
correct decoding of a uniformly distributed $X$ by an informed observer, namely, one that has access
to $Y$, whereas $P_{\mbox{\tiny c}}^{\mbox{\tiny u}}=1/|\calX|$ denotes the probability
of correct blind guessing the value of $X$
(in the absence of $Y$).

\subsubsection{Lempel-Ziv Parsing}
\label{lzp}

The incremental parsing procedure in the LZ78 algorithm is 
a sequential method applied to an input vector $x^n$
over a finite alphabet. In this process, each new phrase is defined as the shortest 
substring that has not appeared previously as a complete parsed phrase, 
except possibly for the final (incomplete) phrase.
For example, applying incremental parsing to the sequence
$x^{15}=\mbox{abbabaabbaaabaa}$ yields
$\mbox{a,b,ba,baa,bb,aa,ab,aa}$. Let $c(x^n)$ designate the
total number of phrases formed from $x^n$ using the incremental parsing procedure
(in the example above, $c(x^{15})=8$).
Also, let $LZ(x^n)$ stand for the length of the LZ78 binary compressed
representation for $x^n$.
By Theorem 2 of \cite{ZL78}, the following inequality
holds:
\begin{eqnarray}
\label{lz-clogc}
LZ(x^n)&\le&[c(x^n)+1]\log\{2\alpha[c(x^n)+1]\}\nonumber\\
&=&c(x^n)\log[c(x^n)+1]+c(x^n)\log(2\alpha)+\nonumber\\
& &\log\{2\alpha[c(x^n)+1]\}\nonumber\\
&=&c(x^n)\log c(x^n)+c(x^n)\log\left[1+\frac{1}{c(x^n)}\right]+
c(x^n)\log(2\alpha)+\log\{2\alpha[c(x^n)+1]\}\nonumber\\
&\le&c(x^n)\log c(x^n)+\log e+
\frac{n(\log \alpha)\log(2\alpha)}{(1-\varepsilon_n)\log
n}+\log[2\alpha(n+1)]\nonumber\\
&\dfn&c(x^n)\log c(x^n)+n\cdot\epsilon(n),
\end{eqnarray}
where we remind that $\alpha$ is the cardinality of $\calX$, and where
$\varepsilon_n$ and $\epsilon(n)$ both tend to zero as $n\to\infty$.
Stated differently, the LZ code-length for $x^n$ is upper bounded by
an expression whose main term is $c(x^n)\log c(x^n)$. On the other hand,
$c(x^n)\log c(x^n)$ is also the dominant term of a lower bound
(see Theorem 1 of \cite{ZL78})
to the shortest code-length attainable by any information lossless finite-state encoder with no
more than $s$ states, provided that $\log(s^2)$ is very small compared to
$\log c(x^n)$. Accordingly, we henceforth refer to $c(x^n)\log
c(x^n)$ as the unnormalized {\em LZ
complexity} of $x^n$ whereas the normalized LZ complexity is defined as
\begin{equation}
\rho_{\mbox{\tiny LZ}}(x^n)\dfn
\frac{c(x^n)\log
c(x^n)}{n}.
\end{equation}

\subsection{Problem Formulation}
\label{pf}

Similarly as in \cite{me13}, we adopt the following model of finite-state
encryption. A finite--state encrypter is defined by a sextuplet
$$E=(\calX,\calY,\calZ,f,g,\Delta),$$
where $\calX$ is a finite input alphabet of size $\alpha = |\calX|$,
$\calY$ is a finite set of variable-length binary strings, including possibly the empty string $\lambda$ (of length zero),
$\calZ$ is a finite set of states,
$f: \calZ \times \calX \times\{0,1\}^* \to \calY$ is the output function,
$g: \calZ \times \calX \to \calZ$ is the next-state function, and
$\Delta: \calZ \times \calX \to {0,1,2,\ldots}$ specifies the number of key bits consumed per step.
When two infinite sequences, $\bx=x_1,x_2,\ldots$, $x_i\in\calX$, henceforth
the {\it plain-text sequence} (or, the source sequence),
and $\bu=u_1,u_2,\ldots$, $u_i\in\{0,1\}$,
$i=1,2,\ldots$, henceforth the {\it key sequence},
are fed into an encrypter $E$, it produces an infinite output sequence
$\by=y_1,y_2,\ldots$, $y_i\in\calY$, henceforth the {\it ciphertext},
while passing through an infinite
sequence of states
$\bz=z_1,z_2,\ldots$, $z_i\in\calZ$, according to
the following recursive equations, implemented for $i=1,2,\ldots$
\begin{eqnarray}
t_i&=&t_{i-1}+\Delta(z_i,x_i),~~~~~~t_0\dfn 0 \label{ti}\\
k_i&=&(u_{t_{i-1}+1},u_{t_{i-1}+2},\ldots,u_{t_i}) \label{ki}\\
y_i&=&f(z_i,x_i,k_i) \label{yi}\\
z_{i+1}&=&g(z_i,x_i) \label{nextstate}
\end{eqnarray}
where the initial state, $z_1$, is assumed fixed, and will be labeled
$z_\star$ hereafter, and where it is understood that if
$\Delta(z_i,x_i)=0$, then $k_i=\lambda$, the null word of
length zero,
namely, no key bits are used in the $i$--th step.
By the same token, if $y_i=\lambda$, no output is produced at this
step, i.e., the system is idling and only the state evolves in response
to the input. In other words, at each time instant $t$, when
the state is $z_i$, the encrypter is fed
by the current plain-text symbol $x_i$ and it consumes the next
$\Delta(z_i,x_i)$ previously unused key bits. It then updates the next state
$z_{i+1}$ and
produces an output $y_i$.

In summary, at each time step $i$:
the current state is $z_i$,
the encrypter receives input $x_i$,
consumes the next $\Delta(z_i, x_i)$ unused key bits from $\bu$ to form $k_i$,
produces output $y_i$, and
transitions to the next state $z_{i+1}$.\\

\noindent
{\bf Remark 1.} Note that the evolution of the state variable $z_i$ depends
solely on the source inputs $\{x_i\}$ and is independent of the key bits. 
This design choice reflects the intended role of $z_i$, which is to retain memory of the source sequence $x^n$, 
allowing the encrypter to exploit empirical correlations and repetitive patterns within the plaintext. 
In contrast, maintaining memory of past key bits—which are assumed to be independent and identically 
distributed (i.i.d.)—offers no practical benefit and is therefore omitted.
That said, the model can be naturally extended to include two state variables: one that evolves based only 
on the source sequence $\{x_i\}$ (as in the current setup), and another that
evolves based on both $\{x_i\}$ and the consumed key bits $\{k_i\}$. In such a framework, the first state variable 
would continue to govern the update of the index $t_i$, while the second could influence the output function, 
allowing for more expressive or adaptive encryption mechanisms.\\

An encrypter with $s$ states, or an $s$--state encrypter,
$E$, is one with $|\calZ|=s$.
It is assumed that the plain-text sequence $\bx$ is deterministic (i.e., an
individual sequence), whereas
the key sequence $\bu$ is purely random, i.e.,
for every positive integer
$n$, $P_{U^n}(u^n)=2^{-n}$.

A few additional notation conventions will be convenient: By
$f(z_i,x_i^j,k_i^j)$, ($i\le j$) we refer
to the vector $y_i^j$ produced by $E$ in response to the inputs $x_i^j$ and
$k_i^j$ when the initial
state is $z_i$. Similarly, the notation $g(z_i,x_i^j)$ will mean the
state $z_{j+1}$ and $\Delta(z_i,x_i^j)$ will designate
$\sum_{\ell=i}^j\Delta(z_\ell,x_\ell)$ under the same circumstances.

As explained in Subsection \ref{bg}, we adopt the maximal leakage of
information as our security metric
given by
\begin{equation}
\calL(x^n\to
Y^n)\dfn\ln\left[\sum_{y^n}\max_{x^n\in\calX^n}P_{Y^n|X^n}(y^n|x^n)\right].\nonumber
\end{equation}
An encryption system $E$ is said to be
{\it perfectly secure} if for every positive integer $n$,
$\calL(x^n\to Y^n)=0$. If 
$\calL(x^n\to Y^n)\to 0$ as $n\to\infty$, we say that the encryption system is asymptotically
secure. 

An encrypter is referred to as {\it information lossless} (IL) if for every
$z_i\in\calZ$, every sufficiently large
$n$ and all pairs
$(x_i^{i+n},k_i^{i+n})$, the quadruple
$(z_i,k_i^{i+n},f(z_i,x_i^{i+n},k_i^{i+n}),g(z_i,x_i^{i+n}))$ uniquely
determines $x_i^{i+n}$.
Given an encrypter $E$ and an input string $x^n$, the encryption key rate of
$x^n$ w.r.t.\ $E$ is defined as
\begin{equation}
\sigma_E(x^n)\dfn \frac{\ell(k^n)}{n}=\frac{1}{n}\sum_{i=1}^n\ell(k_i),
\end{equation}
where $\ell(k_i)=\Delta(z_i,x_i)$ is the length of the binary string $k_i$ and
$\ell(k^n)=\sum_{i=1}^n\ell(k_i)$ is the total length of $k^n$.\\

\noindent
{\bf Remark 2.} It is worth noting that the definition of information losslessness used here
is more relaxed, and thus more general, than the one given in \cite{ZL78}. In \cite{ZL78}, the requirement 
must hold for every positive integer $n$
whereas in the present context, it is only required to hold for all
sufficiently large $n$.
The absence of information losslessness in the stricter sense of \cite{ZL78} does not contradict 
the ability of the legitimate decoder to reconstruct the source. Rather, it implies that reconstructing 
$x^n$  may require more than just the tuple
$(z_i,y_i^{i+n},k_i^{i+n},z_{i+n+1})$, for example,
some additional data from times later than
$i+n+1$ may be needed.\\

The set of all perfectly secure, IL encrypters $\{E\}$ with no more than $s$
states will be denoted by $\calE(s)$.
The minimum of $\sigma_E(x^n)$ over all encrypters in $\calE(s)$
will be denoted by $\sigma_s(x^n)$, i.e.,
\begin{equation}
\sigma_s(x^n)=\min_{E\in \calE(s)}\sigma_E(x^n).
\end{equation}
Finally, let
\begin{equation}
\sigma_s(\bx)=\limsup_{n\to\infty}\sigma_s(x^n),
\end{equation}
and define the {\it finite--state encryptability} of $\bx$ as
\begin{equation}
\sigma(\bx)=\lim_{s\to\infty}\sigma_s(\bx).
\end{equation}
Our purpose is to characterize these quantities and to point out
how they can be achieved in principle.

\section{Main Results}
\label{main}

Our converse theorem, whose proof appears in Section \ref{proof}, is the following.\\

\noindent
{\bf Theorem 1.} For every information lossless
encrypter $E$ with no more than $s$ states,
\begin{eqnarray}
\frac{\calL(x^n\to Y^n)}{n}
&\ge&\bigg[\max_{x^n\in\calX^n}\{\rho_{\mbox{\tiny
LZ}}(x^n)-\sigma_E(x^n)\}-\nonumber\\
& &\delta_s(n)-\frac{(\alpha s-1)\log(n+1)}{n}-\frac{\log s}{n}\bigg]_+,
\end{eqnarray}
where $\delta_s(n)\le O\left(\frac{\log(\log n)}{\log n}\right)$ for every
fixed $s$. Equivalently, 
if $0\le\calL(x^n\to Y^n)\le n\lambda$ for some given constant $\lambda\ge 0$,
then for every $x^n\in\calX^n$ and every information lossless encrypter
$E\in\calE(s)$, 
\begin{equation}
\label{conversebound}
\sigma_E(x^n)\ge\rho_{\mbox{\tiny LZ}}(x^n)-\lambda-\delta_s(n)-
\frac{(\alpha s-1)\log(n+1)}{n}-\frac{\log s}{n}.
\end{equation}

As for achievability, consider first an arbitrary lossless compression scheme that
compresses $x^n$ at a compression ratio of $\rho(x^n)=L(x^n)/n$, and then
applies one-time pad encryption to
$[L(x^n)-n\lambda]_+$ compressed bits. Let $y^n$ denote the resulting
(partially) encrypted compressed
representation of $x^n$. 
Then, obviously, the length of $y^n$, denoted $L(y^n)$, is equal to $L(x^n)$
and so, denoting $L_{\max}=\max_{x^n\in\calX^n}L(x^n)$, we have:
\begin{eqnarray}
\exp_2\{\calL(x^n\to y^n)\}
&=&\sum_{y^n\in\calY^n}\max_{x^n}P_{Y^n|X^n}(y^n|x^n)\nonumber\\
&=&\sum_{\ell=1}^{L_{\max}}\sum_{\{y^n:~L(y^n)=\ell\}}\max_{x^n}P_{Y^n|X^n}(y^n|x^n)\nonumber\\
&=&\sum_{\ell=1}^{L_{\max}}\sum_{\{y^n:~L(y^n)=\ell\}}2^{-[\ell-n\lambda]_+}\nonumber\\
&\le&\sum_{\ell=1}^{L_{\max}}2^{\ell}2^{-[\ell-n\lambda]_+}\nonumber\\
&\le&\sum_{\ell=1}^{L_{\max}}2^{\ell}2^{-(\ell-n\lambda)}\nonumber\\
&=&L_{\max}\cdot 2^{n\lambda},
\end{eqnarray}
and so,
\begin{equation}
\calL(x^n\to y^n)\le n\lambda+\log L_{\max}.
\end{equation}
If $L_{\max}=O(n)$,
then the dominant term is clearly $n\lambda$.\\

\noindent
{\bf Remark 3.} The condition that $L_{\max}=O(n)$ is easy
to satisfy always by a minor modification of any given compression scheme
(if it does not satisfy the condition in the first place). First, test
whether $L(x^n)<\lceil n\log\alpha\rceil$ or
$L(x^n)\ge \lceil n\log\alpha\rceil$. If $L(x^n)<\lceil n\log\alpha\rceil$ add a header bit `0' before
the compressed representation of $x^n$; otherwise, add a header bit `1' and
then the uncompressed binary representation of $x^n$ using
$\lceil n\log\alpha\rceil$ bits. The resulting
code-length would then be $L'(x^n)=\min\{L(x^n),\lceil n\log\alpha \rceil\}+1$
bits.\\

If the compression scheme is chosen to be the LZ78 algorithm then,
\begin{equation}
\sigma_E(x^n)\le \rho_{\mbox{\tiny LZ}}(x^n)-\lambda+O\left(\frac{\log\log
n}{\log n}\right),
\end{equation}
which essentially meets the converse bound (\ref{conversebound}).
We have therefore proved the following direct theorem.\\

\noindent
{\bf Theorem 2.} Given $\lambda\ge 0$, there exists a universal encrypter 
that satisfies
\begin{equation}
\calL(x^n\to Y^n)\le n\lambda +\log n+O(1),
\end{equation}
and for every $x^n\in\calX^n$,
\begin{equation}
\sigma_E(x^n)\le \rho_{\mbox{\tiny LZ}}(x^n)-\lambda+O\left(\frac{\log\log
n}{\log n}\right).
\end{equation}

\noindent
{\em Discussion.}
A few comments are now in order.\\

\noindent
1. We established both a lower bound and an asymptotically matching upper bound on the information leakage, 
leading once again to the conclusion that asymptotically optimal performance can be achieved by applying Lempel-Ziv (LZ) 
compression followed by one-time pad encryption of the compressed bitstream. Together with earlier works 
such as \cite{Ziv78u}, \cite{me13}, and \cite{me23}, this reinforces the message that one-time pad encryption 
applied after LZ compression yields an asymptotically optimal cipher system in several important respects. 
That said, we believe the deeper and more significant contribution of this work lies in the converse theorem (Theorem 1), 
which shows that the key rate required to securely encrypt an individual sequence 
cannot be substantially smaller than its LZ complexity minus the permitted
normalized maximal information leakage.\\

\noindent
2.~Similarly as in \cite{ZL78}, formally there is a certain gap between the converse theorem and
the achievability scheme in its basic form, when examined from the viewpoint of the number of states, $s$, relative to $n$.
While $s$ should be small relative to $n$ for the lower bound to be
essentially $\rho_{\mbox{\tiny LZ}}(x^n)$ (see Subsection \ref{bg} above), the
number of states actually needed to implement LZ78 compression for a
sequence of length $n$ is basically exponential in $n$. In \cite{ZL78}, the
gap is closed in the limit of $s\to\infty$ (after taking the limit
$n\to\infty$) by subdividing the sequence into blocks and restarting the LZ
algorithm at the beginning of every block. A similar comment applies here too in the double
limit of achieving $\sigma(\bx)$.\\

\noindent
3. As discussed in \cite{me23} in a somewhat different context, for an alternative
to the use of the LZ78 algorithm, it can be shown that
asymptotically optimum performance can also be attained by a universal
compression scheme for the class of $k$-th order Markov sources, where $k$ is
chosen sufficiently large. In this case, $\rho_{\mbox{\tiny LZ}}(x^n)$ in
Theorems 1 and 2 should be replaced by the $k$-th order empirical entropy of
order $k$ and some redundancy terms should be modified. But one of these
redundancy terms is $\frac{\log s}{k+1}$, which means that in order to compete
with the best encrypter with $s$ states, $k$ must be chosen significantly
larger than $\log s$, so as to make this term reasonably small.\\

\noindent
4. It is speculated that it may not be difficult extend our findings in
several directions, including: lossy reconstruction,
the presence of side information at either parties, the combination of both,
and successive refinement systems in the spirit of \cite{WBZ25}.
Other potentially interesting extensions are in broadening the scope of the
FSM model to larger classes of machines, including: FSMs with counters,
shift-register machines with counters, and periodically time-varying FSMs with
counters, as was done in Section III of \cite{me23}. Research work in some of these directions is deferred to future studies.

\section{Proof of Theorem 1}
\label{proof}

First, observe that
\begin{equation}
\sigma_E(x^n)=\frac{1}{n}\sum_{i=1}^n\Delta(z_i,x_i)
=\sum_{x,z}\hat{P}(x,z)\Delta(z,x),
\end{equation}
where $\hat{P}=\{\hat{P}(x,z),~x\in\calX,~z\in\calZ\}$ is the joint empirical
distribution of $(x,z)$ derived from $(x^n,z^n)$.
It is therefore seen that $\sigma_E(x^n)$ depends on $x^n$ only via $\hat{P}$.
Accordingly, in the sequel, we will also use the alternative notation $\sigma_E(\hat{P})$ when we
wish to emphasize the dependence on $\hat{P}$.
Let $\calT(x^n)$ denote the set of $\tilde{x}^n\in\calX^n$, that together with their
associated state sequences, share the same empirical PMF
$\hat{P}$ as that of $x^n$ along with its state sequence. Similarly as with $\sigma_E(\cdot)$, we
also denote it by $\calT(\hat{P})$. In the sequel, we will make use of the
inequality
\begin{equation}
\label{LZtype}
\frac{\log|\calT(x^n)|}{n}\ge\rho_{\mbox{\tiny LZ}}(x^n)-\delta_s(n),
\end{equation}
where $\delta_s(n)\to 0$ as $n\to\infty$ for fixed $s$ at the rate of
$\frac{\log(\log n)}{\log n}$. The proof of eq.\ (\ref{LZtype}), which appears in
various forms and variations in earlier papers (see, e.g., \cite{PWZ92}), is
provided in the
appendix for the sake of completeness (see also the related Ziv's inequality
in Lemma 13.5.5 of \cite{CT06}).

For later use, we also define the following sets.
\begin{equation}
\calP(y^n)=\{\hat{P}:~\calT(\hat{P})\cap f^{-1}(y^n)\ne\emptyset\},
\end{equation}
\begin{equation}
\calY(\hat{P})=\{y^n:~\calT(\hat{P})\cap f^{-1}(y^n)\ne\emptyset\},
\end{equation}
where
\begin{eqnarray}
f^{-1}(y^n)&=&\{x^n:~f(z_\star,x^n,k^n)=y^n\nonumber\\
& &~\mbox{for
some}~k^n\in\{0,1\}^{n\sigma_E(x^n)}\}.
\end{eqnarray}
Now, observe that
\begin{eqnarray}
\label{setineq}
|\calY(\hP)|&\ge&\bigg|\bigg\{y^n:~y^n=f(z_\star,x^n,0^{n\sigma_E(x^n)})~\nonumber\\
& &\mbox{for
some}~x^n\in\calT(\hat{P})\bigg\}\bigg|\nonumber\\
&\ge&\max_{z\in\calZ}\bigg|\bigg\{y^n:~y^n=f(z_\star,x^n,0^{n\sigma_E(x^n)})~\nonumber\\
& &\mbox{and}~g(z_\star,x^n)=z~\mbox{for
some}~x^n\in\calT(\hat{P})\bigg\}\bigg|\nonumber\\
&\dfn&\max_{z\in\calZ}|\calY_z(\hP)|\nonumber\\
&\ge&\frac{|\calT(\hat{P})|}{s},
\end{eqnarray}
where the last inequality follows from the following consideration: Let
$x^n$ exhaust all members of $\calT(\hP)$. For each such
$x^n$, let $y^n=f(z_\star,x^n,0^{n\sigma_E(x^n)})$. Now for every $z\in\calZ$, let
$\calT_z(\hP)$ denote the subset of $\calT(\hP)$ for which
$z_{n+1}=g(z_\star,x^n)=z$,
and we have already defined $\calY_z(P)$ to denote the set of corresponding output sequences, $\{y^n\}$.
Obviously, since $\{\calT_z(\hP)\}_{z\in\calZ}$ form a partition of
$\calT(\hP)$, then for some $z=z^*$, $|\calT_{z^*}(\hP)|\ge|\calT(\hP)|/s$, and so,
\begin{eqnarray}
\max_z|\calY_z(\hP)|&\ge&|\calY_{z^*}(\hP)|\nonumber\\
&=&|\calT_{z^*}(\hP)|\nonumber\\
&\ge&\frac{|\calT(\hP)|}{s},
\end{eqnarray}
where the
equality is since the mapping
between $x^n$ and $y^n$ is one-to-one given that $k^n=0^{n\sigma_E(x^n)}$,
$z_1=z_\star$, and $z_{n+1}=z^*$ by the information losslessness postulated,
provided that $n$ is sufficiently large as required.
Now, let $\calY_n^+$ denote that set of all $y^n\in\calY^n$ for which
$P_{Y^n|X^n}(y^n|x^n)>0$ for some $x^n\in\calX^n$. Then,
\begin{eqnarray}
\exp_2\{\calL(x^n\to y^n\}
&=&\sum_{y^n\in\calY_n^+}\max_{x^n} P_{Y^n|X^n}(y^n|x^n)\nonumber\\
&=&\sum_{y^n\in\calY_n^+}\max_{x^n\in\phi^{-1}(y^n)}P_{Y^n|X^n}(y^n|x^n)\nonumber\\
&=&\sum_{y^n\in\calY_n^+}\max_{\hat{P}\in\calP(y^n)}\max_{x^n\in\phi^{-1}(y^n)\cap\calT(\hat{P})}P_{Y^n|X^n}(y^n|x^n)\nonumber\\
&\gea&\sum_{y^n\in\calY_n^+}\max_{\hat{P}\in\calP(y^n)}\max_{x^n\in\phi^{-1}(y^n)\cap\calT(\hat{P})}2^{-n\sigma_E(\hat{P})}\nonumber\\
&=&\sum_{y^n\in\calY_n^+}\max_{\hat{P}\in\calP(y^n)}2^{-n\sigma_E(\hat{P})}\nonumber\\
&\ge&\frac{1}{M_n}\sum_{y^n\in\calY_n^+}\sum_{\hat{P}\in\calP(y^n)}2^{-n\sigma_E(\hat{P})}\nonumber\\
&=&\frac{1}{M_n}\sum_{\hat{P}}\sum_{y^n\in\calY(\hat{P})}2^{-n\sigma_E(\hat{P})}\nonumber\\
&=&\frac{1}{M_n}\sum_{\hat{P}}|\calY(\hat{P})|\cdot 2^{-n\sigma_E(\hat{P})}\nonumber\\
&\geb&\frac{1}{M_ns}\sum_{\hat{P}}|\calT(\hat{P})|\cdot 2^{-n\sigma_E(\hat{P})}\nonumber\\
&\ge&\frac{1}{M_ns}\cdot\max_{x^n\in\calX^n}|\calT(x^n)|\cdot
2^{-n\sigma_E(x^n)}\nonumber\\
&\gec&\frac{1}{M_ns}\cdot\max_{x^n\in\calX^n} 2^{n[\rho_{\mbox{\tiny LZ}}(x^n)-\delta_s(n)]}\cdot
2^{-n\sigma_E(x^n)}\nonumber\\
&=& \exp_2\bigg\{n\cdot\max_{x^n\in\calX^n}\bigg[\rho_{\mbox{\tiny
LZ}}(x^n)-\sigma_E(x^n)-\delta_s(n)-\nonumber\\
& &\frac{\log M_n}{n}-\frac{\log s}{n}\bigg]\bigg\}\nonumber\\
&\ge& \exp_2\bigg\{n\cdot\max_{x^n\in\calX^n}\bigg[\rho_{\mbox{\tiny
LZ}}(x^n)-\sigma_E(x^n)-\delta_s(n)-\nonumber\\
& &\frac{(\alpha s-1)\log(n+1)}{n}-\frac{\log s}{n}\bigg]\bigg\},
\end{eqnarray}
where in (a) we used the fact that $P_{Y^n|X^n}(y^n|x^n)>0$ implies
$P_{Y^n|X^n}(y^n|x^n)\ge 2^{-n\sigma_E(x^n)}$ (because $P_{Y^n|X^n}(y^n|x^n)>0$
implies that there is at least one $k^n\in\{0,1\}^{n\sigma_E(x^n)}$ such that $f(z_\star,x^n,k^n)=y^n$
and the probability of each such $k^n$ is $2^{-n\sigma_E(x^n)}$),
and where $M_n$ is the number of different type classes, $\{\hP\}$, which is upper
bounded by $(n+1)^{\alpha s-1}$. In (b) we used eq.\ (\ref{setineq}) and
in (c) we used eq.\ (\ref{LZtype}).
Finally, the operator $[\cdot]_+$ that
appears in the assertion of Theorem 1 is due to the additional trivial lower
bound $\calL(x^n\to Y^n)\ge 0$.
This completes the proof of Theorem 1.

\section*{Appendix -- Proof of Eq.\ (\ref{LZtype})}
\renewcommand{\theequation}{A.\arabic{equation}}
    \setcounter{equation}{0}

Consider the LZ78 incremental parsing procedure applied to $x^n$ and
let $c_{\ell zz'}$, $\ell\in\calN$, $z,z'\in\calZ$, denote the number of phrases
of length $\ell$, which start at state $z$ and end at state $z'$. Clearly,
$\sum_{\ell,z,z'} c_{\ell zz'}=c(x^n)$, for which we will use the shorthand
notation $c$ in this appendix.

Given that $x^n\in\calT(\hP)$, one can generate other members of $\calT(\hP)$
by permuting phrases of the same length which start at the same state and end at the same state.
Thus, $|\calT(\hP)|\ge\prod_{\ell,z,z'}(c_{\ell zz'}!)$, and so,
\begin{eqnarray}
\log|\calT(\hP)|
&\ge&\sum_{\ell,z,z'}\log(c_{\ell zz'}!)\nonumber\\
&\ge&\sum_{\ell,z,z'}c_{\ell zz'}\log \frac{c_{\ell zz'}}{e}\nonumber\\
&=&\sum_{\ell,z,z'}c_{\ell zz'}\log c_{\ell zz'}-c\log e\nonumber\\
&=&c\sum_{\ell,z,z'}\frac{c_{\ell zz'}}{c}\left[\log\frac{c_{\ell
zz'}}{c}+\log c\right]-c\log e\nonumber\\
&=&c\log c -cH(L,Z,Z')-c\log e,
\end{eqnarray}
where $H(L,Z,Z')$ is the joint entropy of the auxiliary random variables $L$,
$Z$, and $Z'$, jointly distributed according to the distribution
$\pi(\ell,z,z')=c_{\ell zz'}/c$, $\ell\in\calN$, $z,z'\in\calZ$.
To further bound $\log|\calT(\hP)|$ from below, we now derive an upper bound
to $H(L,Z,Z')$:
\begin{eqnarray}
H(L,Z,Z')&\le&H(L)+H(Z)+H(Z')\nonumber\\
&\le&H(L)+2\log s\nonumber\\
&\le&(1+EL)\log(1+EL)-(EL)\log(EL)+2\log s\nonumber\\
&=&\left(1+\frac{n}{c}\right)\log\left(1+\frac{n}{c}\right)-\frac{n}{c}\log\frac{n}{c}+2\log
s\nonumber\\
&=&\frac{n}{c}\log\left(1+\frac{c}{n}\right)+\log\left(\frac{n}{c}+1\right)+2\log
s\nonumber\\
&\le&\log\left(\frac{n}{c}+1\right)+\log(s^2e),
\end{eqnarray}
where the third inequality is due to Lemma 13.5.4 of \cite{CT06}, the
following equality is due to the relation $EL=\sum_{\ell,z,z'}\ell c_{\ell
zz'}/c=n/c$, and the last inequality is due to an application of the
inequality $\log(1+u)\le u\log e$ for all $u>-1$.
It follows that
\begin{eqnarray}
\frac{\log|\calT(\hP)|}{n}
&\ge&\rho_{\mbox{\tiny
LZ}}(x^n)-\frac{c}{n}\log\left(\frac{n}{c}+1\right)-
\frac{c}{n}\log(s^2e)\nonumber\\
&\ge&\rho_{\mbox{\tiny
LZ}}(x^n)-\frac{c}{n}\log\frac{n}{c}-\frac{c}{n}\log\left(1+\frac{c}{n}\right)-
\frac{c}{n}\log(s^2e)\nonumber\\
&\ge&\rho_{\mbox{\tiny
LZ}}(x^n)-\frac{c}{n}\log\frac{n}{c}-\left(\frac{c}{n}\right)^2\log
e-\frac{c}{n}\log(s^2e)\nonumber\\
&=&\rho_{\mbox{\tiny LZ}}(x^n)-\delta_s(n),
\end{eqnarray}
where
\begin{equation}
\delta_s(n)\dfn\frac{c}{n}\log\frac{n}{c}+\left(\frac{c}{n}\right)^2\log
e+\frac{c}{n}\log(s^2e).
\end{equation}
Since $\frac{c}{n}\le\frac{\log\alpha}{\log n}(1+o(1))$ (see eq.\
(6) of \cite{ZL78} and Lemma 13.5.3 of \cite{CT06} and reference therein), the second and the
third terms of $\delta_s(n)$ are bounded by $O(1/\log^2n)$ and $O(1/\log n)$,
respectively. The first term of $\delta_s(n)$ is upper bounded by $O\left(\frac{\log(\log n)}{\log
n}\right)$ (see eq.\ (13.124) of \cite{CT06}).

\end{document}